\documentclass[twocolumn,showpacs,preprintnumbers,amsmath,amssymb]{revtex4}
\usepackage{graphicx}
\usepackage{dcolumn}
\usepackage{bm}
\usepackage{longtable}
\newcommand {\rr }{{\bf r}}

\newcommand {\dd }{{\mbox{d}}}
\newcommand {\be }{\begin{equation}}
\newcommand {\ee }{\end{equation}}
\newcommand {\ba }{\begin{eqnarray}}
\newcommand {\ea }{\end{eqnarray}}

\newcommand {\im}{\mbox{Im}}

\begin{document}

\title{Acoustic attenuation in glasses and
its relation with the boson peak}

\author{W. Schirmacher$^{1}$, G. Ruocco$^{2,3}$, T. Scopigno$^{3}$}

\affiliation{ $^{1}$ Physik-Department E13, Technische Universit\"at
M\"unchen, D-85747, Garching, Germany \\
$^{2}$ Dipartimento di Fisica, Universit\'a di Roma
``La Sapienza'', I-00185, Roma, Italy.\\
$^{3}$ CRS SOFT-INFM-CNR c/o Universit\'a di Roma ``La Sapienza'',
I-00185, Roma, Italy.
}

\begin{abstract}
A theory for the vibrational dynamics in disordered solids
[W. Schirmacher, Europhys. Lett. {\bf 73}, 892 (2006)],
based on the random spatial variation of the shear modulus, has
been applied to determine the wavevector ($k$) dependence of the
Brillouin peak position ($\Omega_k)$ and width ($\Gamma_k$), as
well as the density of vibrational states ($g(\omega)$), in
disordered systems. As a result, we give a firm theoretical ground
to the ubiquitous $k^2$ dependence of $\Gamma_k$ observed in
glasses. Moreover, we derive a quantitative relation between the
excess of the density of states (the boson peak) and $\Gamma_k$, two
quantities that were not considered related before. The successful
comparison of this relation with the outcome of experiments and
numerical simulations gives further support to the theory.
\end{abstract}
\maketitle


The most striking differences between glasses and crystals, at a
macroscopic level, concern the thermal properties. At few tens of
Kelvin, the specific heat of glasses exhibits an excess over the
Debye expectation, which, in a $C(T)/T^3$ {\it vs.} $T$ plot,
appears as a characteristic maximum. Similarly, at low-$T$ the
thermal conductivity increases as $T^2$ with increasing $T$ and,
in the same temperature range where $C(T)/T^3$ has a maximum,
exhibits a plateau. While the first observation can be ascribed to
the presence of an excess of states over the Debye density of
states $g_D(\omega)$  (the ''boson peak'', i.~e. the peak observed
in $g(\omega)/\omega^2$ {\it vs.} $\omega$), the second one
was until recently not understood at all.
Both anomalies appear to be strongly affected
by the characteristics of the normal modes of
vibrations in the $THz$ frequency region. The nature of these
modes has been the subject-matter of a very intense and
controversial debate in the literature since many decades. The
issue has been recently revitalized thanks to new neutron, X-ray,
and other inelastic scattering experiments
\cite{sette,benassi,foret,ruocco99,engberg,ruocco01,ruffle,sascha,mascio,monaco},
computer simulations
\cite{anna,fkaw93,tarask,ribeiro,jund,ruocco00,horbach}, and
analytical theory
\cite{soft,sw93,sdg98,kuhn,gotze,taraskin,bunde,schirm,parisi,turlakov}.

The boson peak shows up in a frequency range where the broadening
of the acoustic excitations becomes of the order of magnitude
of the resonance frequency, thus indicating a possible mode
localization. This observation lead different authors to
hypothesize a relation between the position of the boson peak and
the existence of localized vibrations.
Acoustic waves that become Anderson-localized, indeed, could
produce the plateau in the thermal conductivity.
Following this idea, investigations of (Anderson-)localization
properties of waves in disordered systems based on simulations
\cite{fkaw93}, model calculations \cite{sw93,sdg98} and
field-theoretical techniques \cite{john} have shown that
Anderson-localized states in disordered media do actually occur,
but in a much higher frequency range (near the upper band edge)
than the boson peak frequency.

So the question is still open: what is the very nature of the
states near and above the boson peak frequency? As these states
are neither really propagating nor localized, Feldman and
coworkers \cite{fkaw93} suggested to call them ``diffusons" (they
behave like diffusing light in milky glass). In this regime,
however, the resonance frequency $\Omega_k$ still exhibits a
linear dispersion with the wavevector $k$, albeit with an apparent
sound velocity being somewhat larger than the ultrasonic speed
\cite{ruocco00}. More interesting, in this frequency range the
width $\Gamma_k$ of the excitations, a quantity proportional to
the sound attenuation coefficient, shows a $k^2$ dependence in
most of the investigated materials \cite{ruocco01}, whereas at
lower frequencies the width appears to have a stronger $k$
dependence \cite{mascio}. These findings are still awaiting a
theoretical~explanation.

In this letter we exploit a recently developed theory of
vibrational excitations in disordered elastic media \cite{schirm1}
and apply it to the calculation of the dynamic structure factor.
This theory is based on the model assumption that the disorder
leads to microscopic random spatial fluctuations of the transverse
elastic constant (shear modulus) \cite{sokolov}. As in similar, more schematic
approaches \cite{sdg98,taraskin,schirm} the excess DOS has been shown to
arise from a band of disorder-induced irregular vibrational
states, the onset of which approaches lower frequencies as the
disorder is increased. Within this framework it was possible to
formulate a theory for the energy diffusivity, which gave the
first explanation of the plateau of the temperature dependent
thermal conductivity and its relation to the excess DOS. When
applied to the calculation of the shape of the Brillouin
resonance, this theory {\it explains} {\it i)} the ubiquitous
$k^2$ dependence of $\Gamma_k$ and {\it ii)} the observed
increase of the apparent sound velocity with frequency
\cite{unpublished}. Moreover the theory {\it predicts} {\it iii)}
that the excess of vibrational states, if properly normalized with
the Debye DOS, takes almost a universal value and {\it iv)} the
existence of a quantitative relationship between the excess over
the Debye DOS and the width of the Brillouin line. The latter,
previously unexpected, relation is shown to agree with the
findings of experimental inelastic scattering and computer
simulation experiments.

We now shortly summarize the theory. We consider an elastic medium
with a mass density $m_0$, shear modulus $G$,
bulk modulus 
\mbox{$K=\lambda+\frac{2}{3}G$}
\mbox{($\lambda$ = longitudinal Lam\'e constant).}
These elastic constants are
related to the longitudinal and transverse local sound velocities
as $c_T^2$=$G/m_0$; 
\mbox{$c_L^2$=$(K+\frac{4}{3}G)/m_0$}
\mbox{$=(\lambda+2G)/m_0$.} We now assume
that the shear modulus (and not $\lambda$, which is set to a constant
value $\lambda_0$) exhibits a random spatial variation:
$G(\rr)$=$G_0[1+\Delta\tilde G(\rr)]$. The random function
$\Delta\tilde G(\rr)$ is supposed to be Gaussian distributed
around its average $G_0$ with a variance $\propto \gamma_{G}$. The
parameter $\gamma_{G}$ describes the ``degree of disorder" of the
system. We emphasize that this phenomenological model may be
adequate both for a topologically disordered system (glass) or a
disordered crystal. Using standard field-theoretic techniques
\cite{belitz,john,schirm,schirm1} the {\it self-consistent Born
approximation} (SCBA) for the (complex) self energy function
$\Sigma(\omega)$=$\Sigma'(\omega)+i\Sigma''(\omega)$ has been
derived \cite{schirm1}, resulting in the set of equations:
\begin{eqnarray}
\label{scba}
 \Sigma(\omega)\!\! &=& \!\! \gamma_{G}\Sigma_{k<k_D}
[\chi_L(k,\omega)+\chi_T(k,\omega)] \\
\chi_L(k,\omega)\!\! &=&
k^2[-\omega^2+k^2(c_{L,0}^2-2\Sigma(\omega))]^{-1} \nonumber\\
\chi_T(k,\omega)\!\! &=&
k^2[-\omega^2+k^2(c_{T,0}^2-\Sigma(\omega))]^{-1}. \nonumber
\end{eqnarray}
\begin{figure}
\centerline{\includegraphics[width=8cm,clip=true]{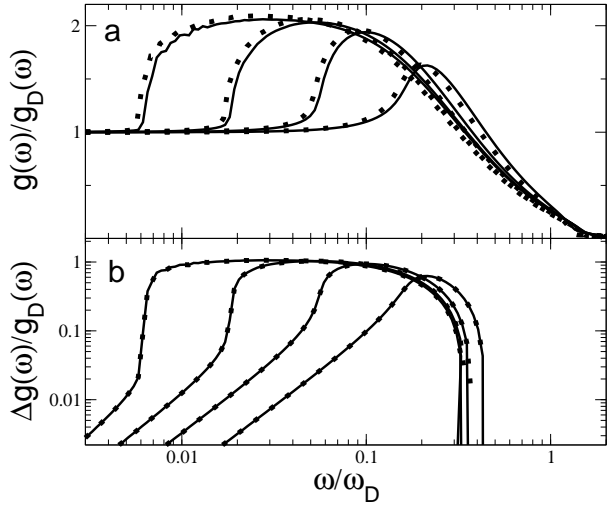}}
\caption{a) Reduced DOS $(g(\omega)/g_D(\omega)$ for
$\epsilon
\!=\!1\!-\!\gamma_G/\gamma_G^c$
 = 0.0001, 0.001, 0.01 and 0.1 (from left to right) and
for $(c_{L}/c_{T})^2=2$ (full lines) and $(c_{L}/c_{T})^2=6$ (dots).
b) Reduced excess DOS $\Delta g(\omega)/g_D(\omega)$ for the same
parameters \cite{note1}.
} \label{fig1}
\end{figure}
Here $\chi_{L}$ ($\chi_{T}$) are the longitudinal and transverse
dynamic susceptibilities, resp., and $c_{L,0}$, $c_{U,0}$ are the
sound velocities of a system without disorder. The sum over $k$
indicates integration up to the Debye cutoff,
$k_D=[6\pi^2N/V]^{\frac{1}{3}}$, according to the identification
$\Sigma_{k<k_D} \rightarrow ({3}/{k_D^3})\int_0^{k_D}k^2\dd k$.
The SCBA is the simplest form of an effective-medium theory for
the disorder \cite{sw93,sdg98,taraskin}, in which the disorder
effects lead to a frequency dependent modification of the sound
velocities (``complex acoustic indices of refraction'') by the
function $\Sigma(\omega)$. The real part $\Sigma'(\omega)$ causes
a renormalization and dispersion of the sound velocities according
to $c_L(\omega)^2=c_{L,0}^2-2\Sigma'(\omega)$ and
$c_T(\omega)^2=c_{T,0}-\Sigma'(\omega)$, whereas its imaginary
part describes the sound attenuation (see below). Apart from the
natural length and frequency scales $k_D^{-1}$ and
$\omega_D=c_Dk_D$ \cite{note1} there are only two non-trivial
parameters in this theory, namely the degree of disorder
$\gamma_{G}$ and the ratio of the renormalized sound velocities
$c_{L}/c_{T}$ \cite{note1}.

From Eq.~(\ref{scba}), the DOS can be calculated as:
\be \label{dos} 
g(\omega)=({2\omega}/{3\pi})
\Sigma_{k<k_D}\frac{1}{k^2}\im\{\chi_L(k,\omega)+2\chi_T(k,\omega)\} \ee
As in other theories of a harmonic solid with quenched disorder
\cite{sdg98,taraskin,parisi} the system becomes unstable if the
disorder exceeds a critical value $\gamma_{G}^c$, which --- in the
present case --- slightly depends on the ratio of sound velocities
and ranges, for example, from $\gamma_G^c$=$0.1666$ for
$c_{L}^2/c_{T}^2$=2 to $\gamma_G^c$=$0.227$ for
$c_{L}^2/c_{T}^2$=6. It has been demonstrated
\cite{sdg98,taraskin,parisi,schirm,schirm1}, that the
quenched disorder produces an excess over the Debye DOS, the
offset of which approaches lower and lower frequencies as
$\gamma_G$$\rightarrow$$\gamma_G^c$.

In Fig.~\ref{fig1} we report the {\it reduced} DOS
$g(\omega)/g_D(\omega)$ (i.~e. the usual boson peak
representation) and the {\it reduced excess} DOS
$[g(\omega)/g_D(\omega)]\!-\!1$ against $\omega/\omega_D$ for
several values of the separation parameter 
$\epsilon
\!=\!1\!-\!\gamma_G/\gamma_G^c$. 
It is remarkable that the value
of the normalized excess does not exceed the value 1, which it
takes for small enough values of $\epsilon$. It is also remarkable
that the excess essentially does not depend on the ratio of the
sound velocities. Moreover, the excess is significantly different
from zero {\it only above} a certain frequency threshold (onset
frequency), below which the DOS basically coincides with the Debye
expectation. The excess DOS turns out to vanish as $\omega^4$ for
$\omega\rightarrow 0$.
\begin{figure}
\centerline{\includegraphics[width=8cm,clip=true]{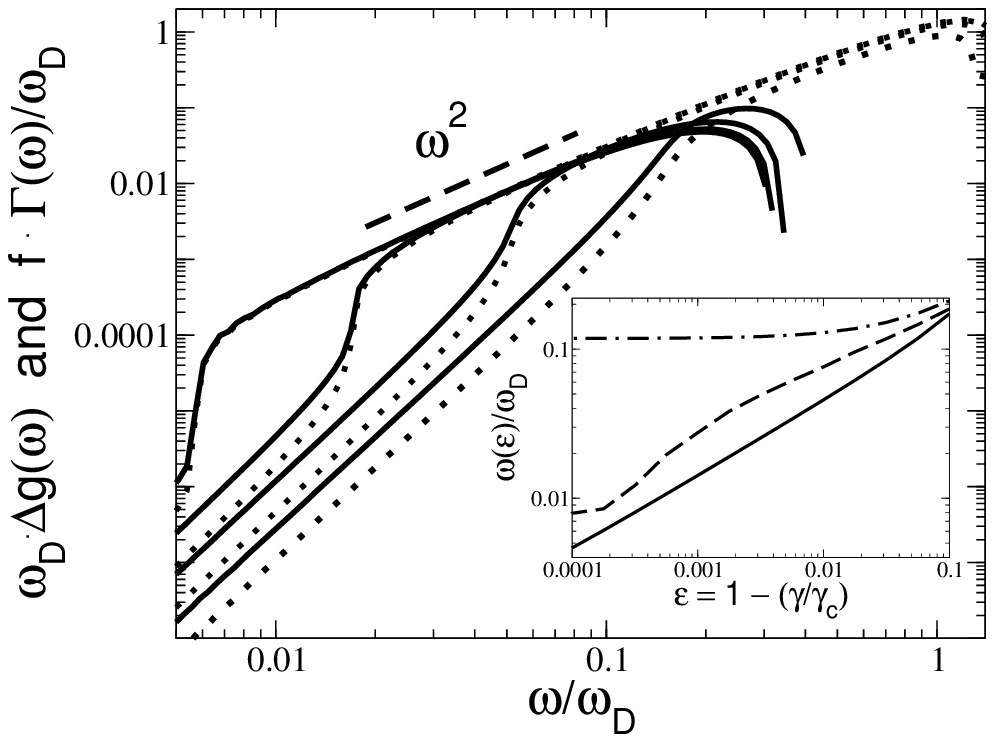}}
\caption{Excess DOS $g(\omega)-g_D(\omega)$ for the same
$\epsilon$ values as in Fig.~\ref{fig1} and for $(c_{L}/c_{T})^2$=6
(full lines) together with the Brillouin linewidth
$f(c_L,c_T)\Gamma(\omega)/\omega_D$=
$2\omega f(c_L,c_T)\Sigma''(\omega)/\omega_D c_L^2$
(dotted lines.)\newline
Insert:
Onset frequency (full line), boson peak frequency (dashed line)
and Ioffe-Regel frequency
$ \omega_{IR}=
c_L(\omega)k/\pi$
(dash-dotted line) as a function of
the parameter
$\epsilon
\!=\!1\!-\!\gamma_G/\gamma_G^c$. 
} \label{fig2}
\end{figure}
\begin{table*}
\begin{ruledtabular}
\begin{tabular}{|l|c|c|c|c|c|c|c|c|c|c|c|c|c|}
Material &$~V/N~$ &$~~~~c_L~~~$ &$~~~~c_T~~~$ &$~~~~A~~~$
&$~~~~B~~~$
&$~~k_D~~$ &$~~c_D~~$
&$~~\omega_D~~$&$~~~A_D~~~$&$~~~\rho~~$ &$~~~R_{Exp}~~$
&$~~~~R_{Th}~~~$&$Err$
\\
\hline
& \AA$^{3}$ & ~~~m/s & ~~~m/s & ps$^{3}$ & ~~~ps &\AA$^{-1}$&m/s&ps&ps$^3$&&ps$^2$&ps$^2$&\%\\
\hline
MD-LJ & 39.5 & 1320 \cite{ruocco00} & ~540 \cite{mazz}    & 2.1 10$^{-2}$ \cite{mazz}    & 0.120 \cite{ruocco00}& 1.14 & 610  & 6.99   & 8.8 10$^{-3}$& 2.4 & 0.099   & 0.103     & 4 \\ 
MD-SiO$_2$   & 14.1 & 5940 \cite{horbach}  & 3570 \cite{horbach} & 3.7 10$^{-5}$ \cite{horbach} & 0.045 \cite{horbach} & 1.61 & 3950 & 63.7   & 1.2 10$^{-5}$& 3.1 & 0.00055 & 0.00060   & 8 \\ 
SiO$_2$      & 15.1 & 5960 \cite{mascio}   & 3750 \cite{mascio}  & 5.3 10$^{-5}$ \cite{buc}     & 0.042 \cite{ruocco99}& 1.58 & 4130 & 65.1   & 1.1 10$^{-5}$& 4.9 & 0.00101 & 0.00053   & 64 \\ 
GeO$_2$      & 16.1 & 3650 \cite{bove}     & 2150 \cite{bove}    & 1.4 10$^{-4}$ \cite{baldi}   & 0.034 \cite{baldi}   & 1.54 & 2380 & 36.8   & 6.0 10$^{-5}$& 2.3 & 0.00237 & 0.00187   & 24 \\ 
Glycerol     & 8.2 & 3600  \cite{scarponi}  & 1870 \cite{scarponi}& 9.9 10$^{-5}$ \cite{wuttke}  & 0.031 \cite{comez02} & 1.93 & 2090 & 40.4   & 4.5 10$^{-5}$& 2.2 & 0.00177 & 0.00194   & 10 \\ 
oTP          & 10.7 & 2940 \cite{tolle}    & 1370 \cite{tolle}   & 3.5 10$^{-4}$ \cite{tolle}   & 0.036 \cite{comez02} & 1.77 & 1540 & 27.3   & 1.5 10$^{-4}$& 2.4 & 0.00445 & 0.00523   & 16 \\ 
Selenium     & 30.0 & 1800 \cite{payen}    & ~895 \cite{payen}   & 2.2 10$^{-3}$ \cite{buc04}   & 0.070 \cite{scopigno}& 1.25 & 1000 & 12.6   & 1.5 10$^{-3}$& 1.4 & 0.00935 & 0.02170   & 80 \\ 
\end{tabular}
\end{ruledtabular}
\caption{
Data taken from the literature together with
quantities derived from them:
atomic volume $V/N$, 
longitudinal and transverse sound velocities $c_{L,T}$;
coefficients $A, A_D$ of the
$\omega^2$ dependence of the measured
and Debye DOS;
coefficient $B$
entering into the frequency dependence of the Brillouin width,
$\Gamma(\omega)$=$B \omega^2$; Debye wavenumber 
$k_D$=$[6\pi^2N/V]^{1/3}$, Debye sound velocity $c_D$ \cite{note1}
and frequency
$\omega_D=c_Dk_D$;
$\rho$
(=$A/A_D$), $R_{exp}$=$(A-A_D)/B$, $R_{Th}$=$f(c_L,c_T)/\omega_D^2$,
$Err$=$|R_{Exp}$-$R_{Th}|
/[(R_{Exp}+R_{Th})/2]$.
}
\end{table*}

We now turn the attention to the dynamical structure factor,
$S(k,\omega)$, which can be measured by inelastic neutron, X-ray
and light scattering:
\ba
&&S(k,\omega)=\frac{1}{\pi}(n(\omega)\!+\!1)\im\{\chi_L(k,\omega)\}=
\frac{1}{\pi}(n(\omega)\!+\!1)\nonumber\\
&&\times{2k^4\Sigma''(\omega) } \left \{ {
[k^2c_L(\omega)^2-\omega^2]^2 +4k^4\Sigma''(\omega)^2 } \right
\}^{-1} \nonumber
\\
&&\approx \frac{1}{\pi}[n(\omega)\!+\!1]\frac{k^2}{2\omega}
\frac{k^2\Sigma''(\omega)/\omega}
{[c_L(\omega)k-\omega]^2+[k^2\Sigma''(\omega)/\omega]^2}
\ea
From this equation we can read off the width of the Brillouin line
(full width at half maximum) as:
\be \label{linewidth1} \Gamma(\omega)
\approx2k^2\Sigma''(\omega)/\omega
\approx\frac{2}{c_L(\omega)^2}\omega\Sigma''(\omega)
\ee
As the ``disorder function'' $\Sigma''(\omega)$ enters also
into the DOS we find from (\ref{dos}) for
$\Sigma''(\omega)<
c_{L,T}(\omega)^2$ the following approximate relation:
\be
\label{deltados}
\omega_D\Delta g(\omega)
=\omega_D[g(\omega)-g_D(\omega)]=f(c_L,c_T)
\,\frac{\Gamma(\omega)}{\omega_D}
\ee

\vspace{-.5cm}

\be\label{factor} \mbox{with}\qquad f(c_L,c_T) =\frac{2}{\pi}
(c_D/c_L)^2
[1+(c_L/c_T)^4] \ee
To check the validity of the approximation made in deriving
Eq.~(\ref{deltados}), in Fig.~\ref{fig2} we compare the excess DOS
with the Brillouin linewidth, multiplied with the factor
$f(c_L,c_T)$. We see that within our theory the approximate
relationship given in Eq.~(\ref{deltados}) holds, particularly in
the regime where both quantities follow approximately an
$\omega^2$ dependence.

Equation (\ref{deltados}) represents the main result of the present
work. It tells us that the Brillouin width $\Gamma(\omega)$ is
proportional to $\omega^2$ (and thus to $k^2$ recalling the
slightly changing value of $c_L(\omega)$) in the whole frequency
region covered by the boson peak, and that it turns to a
$\omega^4$ behavior only below the onset frequency. The onset
frequency, that must not be confused with the boson peak frequency
$\omega_{BP}$ (see insert of Fig. \ref{fig2}), lies well below $\omega_{BP}$, it is located at the
low frequency edge of the boson peak (i.~e. at that frequency
where the excess DOS vanishes) and it tends to zero as
$\epsilon$$\rightarrow$0. Equation (\ref{deltados}) also gives a
quantitative prediction: it establishes a relation between the
absolute value of the excess DOS and the Brillouin peak width. The
material dependent physical quantities entering into this relation
are only macroscopic properties, as sound velocities and density,
that can be determined independently.

In order to test the theoretical prediction manifested in
Equation~(\ref{deltados}) we have collected a number of experimental
and numerical simulation data of the vibrational DOS and the
Brillouin line width, as well as sound velocity data (Table I). A complete
set of data, to our knowledge, only exists in the case of the
numerical simulation of Lennard-Jones particles and for a model of
Silica glass, and for the experimental cases of Silica, Germania,
Selenium, Glycerol and o-Terphenyl.

The first test is that of Fig. 1 a) from which we expect that
the quantity $\rho=\left[g(\omega)/g_D(\omega)\right]_{\rm max}
=A/A_D$ (where $A, A_D$ are the coefficients of the measured
and Debye $\omega^2$ law in the boson peak regime) is equal
to 2.
A systematic variation
upwards is observed, but the overall set of data is not far from
this value \cite{note2}. Exceptions are the case of silica, where the presence
of specific localized modes (tetrahedra rotation) are well
documented, and that of Selenium, where the ratio is low, most
likely because of the boson peak is not well developed, indicating
a large value of $\epsilon$. 

The second test involves 
the quantity
$R_{exp}$=$\Delta g(\omega) / \Gamma(\omega)$=$(A-A_D)/B$, which is
the ratio of the excess DOS to the Brillouin width as determined
by the experiments, while $R_{Th}$ is the same ratio as determined
from the theoretical prediction, i.e.
$R_{Th}$=$f(c_L,c_T)/\omega_D^2$. The values of the two quantities are very
similar in all the cases investigated, the maximum deviation,
observed once more for the cases of silica and selenium, being
within a factor two. The column $Err=\vert R_{Exp}-R_{Th} \vert /
((R_{Exp}+R_{Th})/2)$ reports the percentage of deviation between
the measured and predicted values. Overall, the agreement is very
good, also in view of the approximation made in deriving
Eq.~(\ref{deltados}) and, more important, of the experimental
uncertainties associated with the determination of $A$ and $B$.

In conclusion, we applied  a recent theory of vibrational
excitations in a disordered elastic medium \cite{schirm1} to the
determination of the dynamic structure factor. This theory, that
successfully explained the plateau in the $T$-dependence of the
thermal conductivity in glasses \cite{schirm1}, is now found to
explain one of the most intriguing features in the collective
high-frequency vibrational dynamics of disordered systems: the $k^2$
dependence of the Brillouin linewidth, i.~e. of the acoustic
attenuation. The theory also predicts the existence of a relation
between the excess DOS (the boson peak intensity) and the sound
attenuation coefficient (the Brillouin peak width). A test of this
prediction using existing experimental and simulation data gives
an excellent outcome, thus giving further support to the validity
of the theory itself. Finally, the theory also indicates that the
transition between the $k^4$ law at very low frequency and the
$k^2$ law for $\Gamma_k$ \cite{mascio} (open dots in the insert
of
\ref{fig2}) takes place at a frequency much lower than the boson
peak frequency which, in turn, is much smaller than the
Ioffe-Regel limit frequency.

\end{document}